\documentclass[5p,times]{elsarticle}
\usepackage{ecrc}
\usepackage{amsmath}
\usepackage{booktabs}
\usepackage{array}
\usepackage{tabularx}
\usepackage{dblfloatfix}
\usepackage{float}
\usepackage{placeins}

\newcounter{savedfigure}

\volume{00}

\firstpage{1}

\journalname{Manufacturing Letters}

\runauth{M. Buecheler et al.}

\jid{MFGLET}

\usepackage{amssymb}

\biboptions{numbers,sort&compress}

\usepackage[figuresright]{rotating}

\usepackage[bookmarks=false]{hyperref}
    \hypersetup{colorlinks,
      linkcolor=blue,
      citecolor=blue,
      urlcolor=blue}

\begin{document}
\begin{frontmatter}



\dochead{55th SME North American Manufacturing Research Conference (NAMRC 55, 2027)}%

\title{From Industrial Product Documentation to Automated Digital Product Passport Generation: An Experimental Study of Large Language Models}


\author[a]{Marco Buecheler\corref{cor1}} 
\ead{mbuechel@purdue.edu}
\author[a]{Grant Richards}
\author[a]{Nathan Hartman}

\address[a]{Purdue University, School of Engineering Technology, Knoy Hall of Technology 150, 401 N Grant Street, West Lafayette, IN 47906, USA}

\begin{abstract}
Manufacturing companies that are planning to sell certain products in the European Union (EU) will be required to provide mandatory digital product passports (DPPs). The use of DPPs is to provide standardized product information throughout multiple product life cycles in a circular economy. Required passport information, such as technical specifications, material composition, and environmental product footprint, is available in manufacturers' data sources but is scattered across different heterogeneous systems, documents, and data formats. Manually collecting and structuring the required information for a DPP requires considerable effort, particularly when following standardized information models. The given paper investigates using large language models (LLMs) to automate the generation of DPPs based on the asset administration shell (AAS).

A retrieval-augmented generation (RAG) approach is used to extract product information from industrial documentation and generate populated AAS instances. Product documentation is processed and divided into text chunks. The text chunks are then stored in a persistent ChromaDB vector database using vector embeddings. The retrieval process combines semantic and lexical search to identify relevant information based on standardized AAS submodel template requirements. The retrieved information is provided to five LLMs to generate structured AAS instances. The experiments cover different industrial products and AAS submodels, including `Digital Nameplate', `Technical Data', `Carbon Footprint', and battery-specific passport submodels. Four levels of prompt guidance are investigated to assess their influence on generation quality. 

The generated passports are evaluated against a ground truth. The experimental evaluation includes 4,560 executions across 76 product–submodel combinations, five LLMs, four prompt configurations, and three repetitions per condition. Of the generated passport outputs, 97.4\% are valid JSON, while 44.9\% pass the full AAS validation procedure. Under deterministic scoring, mean completeness, conditional accuracy, and overall field correctness (OFC) are 49.6\%, 58.3\%, and 30.8\%, respectively. An LLM-as-a-judge yields better scores of 66.5\%, 88.9\%, and 60.4\% for the same outputs. Deterministic OFC increases from 13.4\% with `Low' prompt guidance to 30.2\% when the target AAS template is provided and approximately 40\% with additional guidance.
\end{abstract}

\begin{keyword}
Digital Product Passport \sep Digital Battery Passport \sep Asset Administration Shell \sep Retrieval-Augmented Generation \sep Large Language Models \sep Circular Economy
\end{keyword}

\cortext[cor1]{Corresponding author. Tel.: +1 (765) 494-9099.}

\end{frontmatter}



\section{Introduction} \label{introduction}
Circular manufacturing aims to retain the value of products, individual components, and materials \cite{morseletto_targets_2020}. Decisions throughout these activities depend on the availability of product information. Technical specifications may determine whether a product's components can be reused elsewhere. Information on material composition is useful in disassembly and recycling processes. Environmental data, including product carbon footprint, can be used to assess the environmental implications of product life cycle activities. However, the needed product information is often scattered across many heterogeneous data systems in an organization. Moreover, the data originates from different activities and stakeholders in a company's value chain. 

\newpage
\begin{nomenclature}
\begin{deflist}[XXXX]
\defitem{AAS}\defterm{Asset Administration Shell}
\defitem{CAD}\defterm{Computer-Aided Design}
\defitem{DBP}\defterm{Digital Battery Passport}
\defitem{DPP}\defterm{Digital Product Passport}
\defitem{DSR}\defterm{Design Science Research}
\defitem{ESPR}\defterm{Ecodesign for Sustainable Products Regulation}
\defitem{EU}\defterm{European Union}
\defitem{IDTA}\defterm{Industrial Digital Twin Association}
\defitem{LLM}\defterm{Large Language Model}
\defitem{OCR}\defterm{Optical Character Recognition}
\defitem{OFC}\defterm{Overall Field Correctness} 
\defitem{RAG}\defterm{Retrieval-Augmented Generation}
\defitem{RQ}\defterm{Research Question}
\defitem{RRF}\defterm{Reciprocal Rank Fusion}
\end{deflist}
\end{nomenclature} 

To streamline the data for circular economy purposes, the European Union (EU) will require companies selling products in the EU to provide digital product passports (DPPs). These DPPs are product records that offer an approach to make product data available throughout a product's (multiple) life cycle(s). The EU has introduced a regulatory framework for DPPs through the `Ecodesign for Sustainable Products Regulation (ESPR)' \cite{european_commission_regulation_2024}. The regulation establishes a framework for product-specific information requirements intended to support environmental sustainability and circularity. 

The `EU Batteries Regulation' establishes digital battery passport (DBP) requirements for industrial batteries \cite{european_commission_regulation_2023}. Depending on the product and applicable requirements, passports may include information on product identification, technical characteristics, material composition, environmental impacts, and other life cycle-related information \cite{battery_pass_consortium_battery_2023, european_commission_guidance_2026, jansen_stop_2023}.

To meet the information requirements of a DPP or a DBP, respectively, the required product documentation often needs to be extracted from heterogeneous data sources. For many companies, this process can be time-consuming and requires manual effort as well as deep product knowledge.\\

Recent advances in large language models (LLMs) offer opportunities to automate the extraction and structuring of product information required for DPPs. Previous research highlights the ability of LLMs to interpret unstructured data and extract semantically relevant information \cite{chen_can_2025, menon_automated_2025, xia_generation_2024}. However, there are limited quantitative results on the performance of LLMs in the field. Therefore, the given study evaluates the automated creation of DPPs with five different LLMs and multiple industrial product documents from different manufacturers. Eight product categories are part of the experimental design. To standardize the passport creation process, the LLMs are prompted to create DPPs based on the data structure of an asset administration shell (AAS). The study uses retrieval-augmented generation (RAG) to identify relevant product information in a vector database, with retrieval queries derived from standardized AAS submodel requirements. The experiments cover different standardized AAS submodels, including `Digital Nameplate', `Technical Data', `Carbon Footprint', and also battery-specific passport submodels \cite{industrial_digital_twin_association_ev_registered_2026}. Four levels of prompt guidance are investigated to determine their influence on the generation results. The generated AAS instances are evaluated for structural validity, information completeness, conditional accuracy, overall field correctness (OFC), and consistency across repeated runs. Particular attention is given to differences between submodels that require technical, environmental, and material-related information.

As a result, the study wants to answer the following four research questions (RQs) quantitatively:
\begin{itemize}
    \item \textit{RQ1:} To what extent can LLMs generate complete, accurate, and structurally valid AAS instances from heterogeneous industrial product documentation using requirement-driven RAG for automated DPP generation?
    \item \textit{RQ2:} How do LLM selection and prompt guidance affect the quality of generated DPPs?
    \item \textit{RQ3:} How does generation performance differ across industrial product categories and standardized AAS submodels?
    \item \textit{RQ4:} How consistent are the generated AAS instances across repeated runs under identical experimental conditions?
\end{itemize}
 
The following Section \ref{sec:background} outlines how the current study relates to already existing work and adds to the current knowledge of research. Section \ref{sec:methodology} focuses on the methodology of the given automated DPP generation pipeline, and Section \ref{sec:exp-design} outlines the experimental design of the LLM evaluation study. The exact evaluation metrics that are being evaluated in each of the RQs are explained in Section \ref{sec:metrics}. The findings of the study are pointed out in Section \ref{sec:results}. The last two Sections \ref{sec:discussion} and \ref{sec:conclusion}, discuss and conclude the findings and outline needs for future research activities.

\section{Background and Related Work}\label{sec:background}
Multiple research fields need to be considered for the given paper. Therefore, the background and related work section outlines the state-of-the-art regarding DPP information exchange requirements for circular manufacturing, data representations for DPPs, and LLM-based DPP/AAS generation approaches (including the storage and retrieval of product evidence). Lastly, identified research gaps are summarized.

\subsection{DPP Interoperability Needs for Circular Manufacturing}
Product passports are digital data records that consolidate product-relevant information generated across the entire value chain, from raw material extraction through manufacturing, use, and end-of-life \cite{adisorn_towards_2021, walden_digital_2021, jansen_current_2022}.
Research on DPP systems identifies the collection and exchange of product life cycle information as a central implementation challenge \cite{jansen_stop_2023}. Industrial batteries are the first product category where a passport will be required \cite{european_commission_regulation_2023}. The `Battery Pass Consortium's' content guide translates battery-passport requirements into practical information categories. These contributions establish that passport preparation involves information from different activities and stakeholders rather than a single product document \cite{battery_pass_consortium_battery_2023}. 

\subsection{Standardized Data Representations}
The AAS provides a standardized approach for representing industrial assets \cite{alexopoulos_why_2025}. An AAS identifies the represented asset and references submodels that describe specific information about that asset. The Industrial Digital Twin Association (IDTA) publishes standardized AAS submodel templates to support interoperability between different organizations and systems. Each submodel contains elements such as properties, collections, and references \cite{industrial_digital_twin_association_ev_specification_2026}. As a result, a product's information is represented in a structured and machine-readable form. 

The information requirements and structures differ across submodels. Product identification and manufacturer information are recorded in the `Digital Nameplate'. The `Technical Data' submodel describes technical product characteristics(including standardized property definitions and units). Environmental impact information is found in the `Carbon Footprint' submodel. For batteries, the `Battery Material Composition' submodel informs about battery materials and substances.

The following Section \ref{sec:AAS-generation} reviews previous research on the automated generation of AAS submodels using LLMs.

\subsection{AAS generation with LLMs}\label{sec:AAS-generation}
Xia et al. \cite{xia_automated_2022} examine automated mapping between heterogeneous information models and standardized AAS representations using semantic fingerprints. Their approach uses language-model-based semantic representations to identify relationships between source properties and standardized information elements.

In subsequent work, Xia et al. \cite{xia_generation_2024} explore AAS generation using LLM agents and compare different models and a retrieval-augmentation mechanism for ECLASS selection. Their work establishes LLM-based transformation of technical documentation into digital-twin representations.\\

Shi et al. \cite{shi_dual_2025} address the mapping of information between existing industrial data systems and AAS representations. The work uses semantic matching between source information and standardized AAS property definitions. Additionally, the use of classification systems such as ECLASS is included. The approach is relevant to enhancing information interoperability, as equivalent technical properties are often described using different terminology in different stakeholders' systems.\\

More recently, B\"ucheler et al. \cite{bucheler_regulation_2026} investigate the generation of regulatory compliance artifacts using LLMs, focusing on DBPs under the `ESPR' and `EU Batteries Regulation'. Five LLMs are evaluated using different levels of prompt guidance and multiple repeated executions per experimental condition. For battery passports, the study uses an AAS `Battery Material Composition' submodel to evaluate the completeness and consistency of generated information. The results show high levels of required-field inclusion and consistency for most models, with relatively small differences between prompt configurations.\\ 

Groß and Heidrich \cite{gros_aas-rail_2026} propose AAS-RAIL, which uses previously modeled AAS instances to generate extraction guidance for retrieving technical property values from product datasheets. AAS-RAIL retrieves property-specific extraction helpers generated from existing AAS instances and their associated datasheets. These helpers describe how values should be identified and formatted. A benefit is that manufacturer-specific conventions are taken into account without having to fine-tune the LLM. Their evaluation focuses on the `Technical Data' submodel extraction and reports improvements over a fixed few-shot baseline. A downside of the approach is that constructing the retrieval requires previously modeled AAS examples \cite{gros_aas-rail_2026}.\\

Vogel et al. \cite{vogel_extraction_2025} evaluate the extraction of technical product information from manufacturer datasheets using LLMs and standardized property definitions. The authors distinguish between extracting available product properties followed by semantic annotation and identifying values based on predefined property definitions. The authors use the open-source \texttt{basyx-pdf-to-aas} library to extract technical properties from manufacturer datasheets. They base their extraction on predefined property definitions, such as those from ECLASS. The extracted values are compared with the information contained in previously populated AAS `Technical Data' submodels from several manufacturers to evaluate extraction accuracy.

\subsection{Summary of Identified Research Gaps}
Previous research has demonstrated the feasibility of using LLMs for technical information extraction and AAS generation. However, the extent to which requirement-driven RAG can generate complete, factually correct, and structurally valid AAS instances from heterogeneous industrial product documentation remains insufficiently investigated. Existing approaches primarily focus on technical submodels. However, DPP applications require information from various domains such as product identification, material composition, and environmental information. Consequently, further contributions must be distinguished by their product data retrieval mechanisms rather than a general claim to introduce automated AAS generation. This motivates the evaluation of requirement-driven RAG for automated DPP generation effectiveness across multiple standardized AAS submodels and industrial product categories (\textit{RQ1}).\\

\begin{figure*}[!b]
\centering
\includegraphics[width=\textwidth]{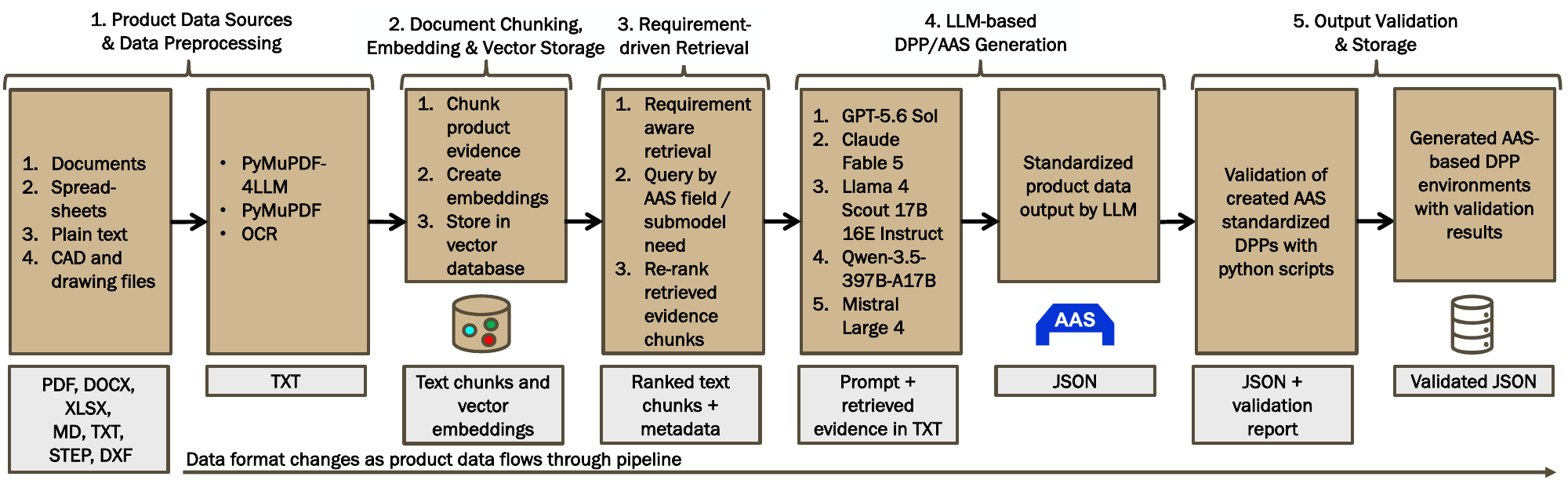}
\caption{Overview of the proposed DPP/AAS generation pipeline.}
\label{fig:pipeline}
\end{figure*}

Furthermore, the influence of LLM selection and prompt guidance on generation quality across different AAS submodels has not been systematically established. Although previous studies have compared LLMs and investigated prompting strategies, it remains unclear whether their findings extend to submodels with different information requirements and structures across product categories. The present study therefore examines differences across LLM prompt configurations (\textit{RQ2}), as well as variations in generation performance across product categories and standardized submodels (\textit{RQ3}).\\

Finally, the reproducibility of generated AAS instances requires further investigation. Previous evaluations have considered output consistency, but limited evidence is available on the stability of field presence and generated values across repeated executions of requirement-driven RAG for different AAS submodels. The study therefore evaluates cross-run consistency under identical experimental conditions (\textit{RQ4}).\\

How the present study investigates the automated generation of AAS-based DPP information across heterogeneous industrial products and standardized submodels is explained in further detail in the following Section \ref{sec:methodology}.

\section{Methodology}\label{sec:methodology}
The study follows a design science research (DSR) approach to guide the development and evaluation of a requirement-driven RAG pipeline for automated DPP creation. In DSR, practical problems are addressed through the systematic development and evaluation of an artifact \cite{hevner_three_2007}. The artifact is based on the current state of knowledge while also contributing to new scientific results itself \cite{hevner_three_2007, hevner_design_2004}.\\

The proposed approach transforms industrial product documentation into structured AAS representations for automated DPP creation. It combines document ingestion, requirement-driven retrieval (based on the specific AAS submodel under investigation), standardized AAS-based passport generation through an LLM, as well as a final output validation check. The generated output is an `AAS 3.2 Environment' serialized in JSON format. Each generation task is defined by a selected product, its associated source documentation, and a target standardized AAS submodel. The submodel determines both the information required for retrieval and the preferred structure of the generated output. \textit{Figure} \ref{fig:pipeline} shows the individual components of the pipeline and how product data flows through the pipeline. 

\subsection{Product Data Sources and Data Preprocessing}
 The first stage of the pipeline preprocesses industrial documents and converts the data content into text suitable for information retrieval and LLM processing. The implementation includes document readers for supported file formats, including PDF and Word documents, spreadsheets, text and Markdown files, or specialized formats such as computer-aided design (CAD) files or drawings. 
 
The tools PyMuPDF4LLM and PyMuPDF are primarily used to process documents with extractable text. Optical Character Recognition (OCR) is available as a fallback when the document does not contain sufficient extractable text. The goal is to preserve relevant product information as written in the source documents. Such information includes technical properties, numerical values, units, and other product descriptions.
 
For each experimental condition, the source documents are assigned to the corresponding product. 

\subsection{Document Chunking, Embedding, and Vector Storage}
Following preprocessing, the extracted document text is divided into chunks with a target length of 1,400 characters and an overlap of 200 characters to preserve information spanning chunk boundaries. Each chunk is converted into a vector embedding using the sentence-transformers model \texttt{all-MiniLM-L6-v2}. The resulting embeddings are stored in a persistent ChromaDB vector database. The vector database is configured for cosine-distance similarity retrieval. Each database record contains the original text chunk and metadata identifying its source document and chunk position.

The persistent database enables the indexed documentation to be reused across experimental executions without repeating the embedding process for each LLM request. Retrieval is restricted to the source documents assigned to the current product to prevent the inclusion of unrelated product information. To support the comparability of the experiments, the same indexed documentation and retrieval configuration are maintained across LLMs and prompt conditions. If inputs exceed \texttt{all-MiniLM-L6-v2's} default limit of 256 word pieces, input truncation occurs. Some chunks may exceed the limit as chunk boundaries are defined in characters rather than tokens. As a result, information is potentially excluded near the end of the vector representation. The complete chunk text remains stored in ChromaDB and accessible to lexical retrieval. The potential impact of embedding truncation on retrieval performance will also be evaluated by examining the retrieved evidence included in the prompts provided to the LLMs.

\subsection{Requirement-Driven Retrieval}
The retrieval process begins with the standardized AAS submodel selected for the experiment. The pipeline identifies relevant submodel elements and uses their identifiers to generate retrieval queries. The identifiers are converted into readable field descriptions and combined with product-specific information. For each selected field, the pipeline generates queries that include the product designation and the corresponding AAS field name, together with field-specific queries that do not repeat the product designation. Additional queries provide general product and submodel information.
The retrieval process combines dense semantic search using the ChromaDB embeddings with lexical retrieval based on BM25. Both methods produce candidate document chunks that may contain information relevant to the selected submodel requirement. The candidate rankings are combined using reciprocal rank fusion (RRF). This method assigns a combined ranking score based on the position of each chunk in the individual retrieval lists.  After rank fusion, a cross-encoder reranking model, ms-marco-MiniLM-L6-v2, evaluates the relevance of candidate chunks to the corresponding requirement. Unlike vector similarity search, the cross-encoder processes the query and candidate text together to produce a relevance score.

The retrieval configuration selects one main evidence chunk for each requirement when a suitable candidate is available. To preserve nearby information (such as table headings, units, and explanatory text), preceding and following chunks are also included. Duplicate chunks are removed when constructing the final retrieved evidence.
The retrieval process follows a requirement-driven strategy. It does not select a fixed global number of chunks for the complete submodel. Each submodel requirement is used to retrieve relevant information. This is intended to reduce the risk that frequently mentioned product characteristics dominate the retrieved evidence while less frequently documented properties are omitted. For applicable `Technical Data' conditions, the implementation also incorporates ECLASS-related information to support standardized technical property identification.

\subsection{LLM-based DPP/AAS Generation}
Following retrieval, the selected document chunks are provided to the LLM together with the generation instructions assigned to the experimental condition. The generation task requires the model to extract product-specific information and create a populated AAS Environment in JSON format. The instructions specify that the generated environment must contain the AAS and the target submodel, and that the AAS must reference the generated submodel. The models are instructed to use only values supported by the provided product documentation and not to invent missing information. 

The information supplied to the model and the amount of generation guidance vary according to the assigned prompt configuration. The five investigated LLMs and four prompt configurations are described in Section \ref{sec:exp-design}.

\subsection{Output Validation and Storage}
The generated AAS environments are evaluated through separate validation and scoring procedures. The first stage determines whether the generated response can be parsed as a valid JSON file. Successful parsing establishes syntactic validity but does not establish compliance with the AAS metamodel. Therefore, a subsequent validation step evaluates the generated passport environment against an AAS 3.2 JSON schema and also performs additional structural checks. These checks include the consistency of AAS elements, references between the AAS and its submodel, and uniqueness requirements for applicable identifiers. The generated outputs are retained together with their execution metadata and validation results. Field-level quality is evaluated using two scoring approaches applied to the same 4,560 generated outputs. The first approach is a deterministic rule-based comparison against the reference AAS environments, and the second approach is an LLM-as-a-judge that reports evaluated field presence and correctness separately. The two evaluation approaches allow analysis of differences in scoring outcomes without changing the generated outputs. For malformed JSON responses containing recoverable field information, the recovered fields are included in field-level scoring, while JSON and AAS validity are assessed on the original outputs. The exact evaluation procedure is explained in Section \ref{sec:metrics}.

\section{Experimental Design}\label{sec:exp-design}
\subsection{Product Dataset and AAS Submodels}
The experimental dataset consists of industrial product documentation associated with different product categories and manufacturers. A total of 35 products are selected, resulting in 76 product–submodel combinations across eight product categories. The investigated products include industrial batteries, diesel engines, unmanned aerial vehicles, electrical components, and electric motors. The product data is ingested in different document formats with different structures and levels of information content (see Table \ref{tab:data-formats} on Page \pageref{tab:data-formats}).

\begin{table}[!ht]
\centering
\caption{Data formats tested in the pipeline}
\begin{tabular}{ll}
\toprule
\textbf{Category} & \textbf{Formats} \\
\midrule
Documents               & PDF, DOCX \\
Spreadsheets            & XLSX \\
Plain text              & MD, TXT \\
CAD and drawing files   & STEP, DXF \\
\bottomrule
\end{tabular}
\label{tab:data-formats}
\end{table}

Different standardized submodels are selected according to the available product information and the intended passport representation. The experiments include general `Digital Nameplate' and `Technical Data' submodels, `Carbon Footprint' submodels, and battery-specific submodels covering battery nameplate information, technical characteristics, product carbon footprint, and material composition.

Each product–submodel condition is associated with an experiment definition identifying the product, the permitted source documents, and the target AAS template. Table \ref{tab:categories-submodels} gives an overview of product categories, tested submodels, number of product-submodel
conditions, and resulting generation runs.

\begin{table}[!ht]
\centering
\caption{Overview of experimental conditions}
\footnotesize
\begin{tabularx}{\columnwidth}{>{\raggedright\arraybackslash}p{2.2cm} >{\raggedright\arraybackslash}X >{\centering\arraybackslash}p{1.1cm} >{\centering\arraybackslash}p{1.1cm}}
\toprule
\textbf{Product Category} & \textbf{Submodels Tested} & \textbf{Conditions} & \textbf{Runs} \\
\midrule
Battery                 & Battery Digital Nameplate, Battery Technical Data, Battery Material Composition, Battery Product Carbon Footprint & 7  & 420 \\
Car                     & Digital Nameplate, Carbon Footprint & 2  & 120 \\
Electric Motor          & Digital Nameplate, Technical Data & 2  & 120 \\
Electrical Component    & Digital Nameplate, Technical Data, Carbon Footprint & 44 & 2,640 \\
Engine                  & Digital Nameplate, Technical Data, Carbon Footprint & 13 & 780 \\
Flight Controller       & Digital Nameplate, Technical Data & 2  & 120 \\
Unmanned Aerial Vehicle & Digital Nameplate, Technical Data & 4  & 240 \\
Ventilation Unit        & Digital Nameplate, Carbon Footprint & 2  & 120 \\
\midrule
\textbf{Total}          &  & \textbf{76} & \textbf{4,560} \\
\bottomrule
\end{tabularx}
\label{tab:categories-submodels}
\end{table}

\subsection{Experimental Factors}
Five LLMs are evaluated using the outlined requirement-driven RAG architecture. The selected LLMs include closed and open-weight models and are shown in Table \ref{tab:models}.

\begin{table}[!ht]
\centering
\caption{Overview of evaluated LLMs}
\begin{tabular}{lcc}
\toprule
\textbf{Name} & \textbf{Developer} & \textbf{Type} \\
\midrule
GPT-5.6 Sol          & OpenAI        & Closed \\
Claude Fable 5       & Anthropic     & Closed \\
Llama 4 Scout        & Meta          & Open   \\
Qwen3.5-397B-A17B    & Alibaba       & Open   \\
Mistral Large 4      & Mistral AI    & Open   \\
\bottomrule
\end{tabular}
\label{tab:models}
\end{table} 

The experiments investigate the influence of LLM selection and prompt guidance on the generated AAS instances. The four prompt configurations, `Low', `Baseline', `Medium', and `High', are applied to each model. Table \ref{tab:prompts} provides an overview of the four prompt configurations that are being investigated.

\begin{table}[!ht]
\centering
\caption{Context levels provided to the model in the generation task.}
\label{tab:prompts}
\small
\begin{tabular}{l p{0.72\columnwidth}}
\toprule
\textbf{Level} & \textbf{Information provided to the model} \\
\midrule
Low & Generation task, product information, and retrieved document evidence. The full target submodel template is not provided. \\
Baseline & Low-level inputs plus the target standardized AAS submodel template. \\
Medium & Baseline inputs plus the template's accompanying documentation. \\
High & Medium-level inputs plus a written summary of the AAS metamodel. \\
\bottomrule
\end{tabular}
\end{table}

The full model and prompt configurations can be found in the GitHub repository, which is linked in \ref{sec:materials}. 

Each experimental condition is executed three times to examine variability across repeated generations. The planned experiment therefore comprises the product–submodel conditions, five models, four prompt configurations, and three repeated runs per condition. For the 76 registered product-submodel conditions, this corresponds to 4,560  generation attempts if all combinations are completed, as Eq. \eqref{eq:total_runs} shows.

\vspace{-5mm}
\begin{equation}
N = N_{\mathrm{exp}} \times N_{\mathrm{LLM}} \times N_{\mathrm{prompt}} \times N_{\mathrm{rep}} = 76 \times 5 \times 4 \times 3 = 4{,}560
\label{eq:total_runs}
\end{equation}

\subsection{Ground Truth Construction}
The generated AAS instances are compared against manually prepared ground truth references. Each reference contains the expected AAS and one target submodel populated using the corresponding product documentation. The reference data are associated with source-supported values and the relevant AAS information structure. Required AAS elements are retained (even without a source-supported value) according to the cardinality constraints of the corresponding IDTA templates.

The evaluation distinguishes fields for which a source-supported value is available from fields for which the reference explicitly expects an empty value. This distinction is important because a missing source value should not be replaced with an invented value to increase the apparent completeness of the generated submodel.

The ground-truth environments are used to evaluate the presence, structure, and correctness of individual generated fields. The exact scoring procedure is explained in Section \ref{sec:metrics}.

\subsection{Experimental Procedure}
For each experimental condition, the applicable product documentation is retrieved from the persistent vector database using queries derived from the selected AAS submodel requirements (as explained in Section \ref{sec:methodology}).

The retrieved evidence is combined with the assigned prompt configuration and submitted to the selected LLM. The generation process is repeated three times using the same experimental settings.
Each response is saved together with its execution metadata, including the selected model, prompt configuration, and product–submodel condition. After generation, the responses are processed by the validation and scoring procedures.

\section{Evaluation Metrics}\label{sec:metrics}
The generated AAS environments are evaluated for output validity, information completeness, conditional accuracy, OFC, and consistency across repeated executions. The two scoring approaches used in the study are explained in the following Subsection.

\subsection{Deterministic and LLM-as-a-Judge Evaluation}
Two independent scoring approaches are used to evaluate the generated AAS instances against the manually constructed ground-truth environments.\\ 

The first approach, deterministic scoring, relies on predefined rules to match generated fields with the corresponding reference fields and assess their correctness. Field presence is determined through structural matching. Value correctness is evaluated using rule-based comparisons of normalized representations, including text, numerical values, dates, or units. These comparisons follow predefined criteria.\\ 

The second approach, LLM-as-a-judge, uses a separate language model (OpenAI's GPT-6.1 Terra) to evaluate the generated fields against the ground-truth reference. The judge independently determines whether each required field is present and whether its generated value is correct. This approach can potentially recognize semantically equivalent information that is expressed differently from the reference. However, it may also introduce variability or more permissive judgments \cite{thakur_judging_2025}. The judge's decisions are independent of the deterministic scoring results. 
Comparing the two approaches therefore provides insight into the influence of the evaluation method on the measured quality of automatically generated AAS instances. Both approaches use the same required reference fields and the same mathematical definitions of completeness, conditional accuracy, and OFC, presented below. 

\subsection{Completeness Scoring}
Completeness evaluates the proportion of required submodel fields populated in a generated AAS instance. For a single execution, completeness $C$ is calculated as

\vspace{-5mm}
\begin{equation}
\label{eq:completeness}
C = \frac{1}{F}\sum_{i=1}^{F}\delta(\hat{v}_i),
\end{equation}

\noindent where $F$ is the number of required fields, $\hat{v}_i$ is the generated value of field $i$, and

\vspace{-5mm}
\begin{equation}
\label{eq:population-indicator}
\delta(\hat{v}_i) =
\begin{cases}
1, & \text{if field $i$ is populated with a non-null value,}\\
0, & \text{otherwise.}
\end{cases}
\end{equation}

A populated value contributes to completeness irrespective of whether the
value is factually correct. Therefore, completeness and accuracy are reported separately because a populated field may contain an incorrect value.

\subsection{Conditional Accuracy Scoring}
Conditional accuracy assesses the correctness of populated fields against the ground truth. Comparisons are carried out after appropriate normalization of text, dates, numeric formats, and units. For a single execution, conditional accuracy $A$ is defined as

\vspace{-5mm}
\begin{equation}
\label{eq:conditional-accuracy}
A = \frac{1}{F_A}\sum_{i=1}^{F_A}\delta(\hat{v}_i,v_i^*),
\end{equation}

\noindent where $F_A$ denotes the number of ground-truth fields populated in the generated AAS, $v_i^*$ is the reference value, and

\vspace{-5mm}
\begin{equation}
\label{eq:correctness-indicator}
\delta(\hat{v}_i,v_i^*) =
\begin{cases}
1, & \text{if the comparison criterion is met,}\\
0, & \text{otherwise.}
\end{cases}
\end{equation}

All populated ground-truth fields are included in this calculation. Fields for which correctness cannot be established receive no correctness credit. Conditional accuracy is undefined if $F_A=0$.

\subsection{Overall Field Correctness Scoring}
The OFC measures the proportion of all required fields that are both populated and correctly extracted. Missing or incorrectly populated fields are counted as failures:

\vspace{-5mm}
\begin{equation}
\label{eq:ofc}
OFC = \frac{1}{F}\sum_{i=1}^{F}\delta(\hat{v}_i,v_i^*),
\end{equation}

\noindent where the indicator in Eq. \eqref{eq:ofc} is $1$ only if the required field is populated and its value satisfies the ground-truth comparison criterion.
When completeness and conditional accuracy refer to the same set of required fields,

\vspace{-5mm}
\begin{equation}
\label{eq:ofc-product}
OFC = C\times A.
\end{equation}

Thus, OFC combines completeness and conditional accuracy in one measure.

\subsection{Consistency Across Repeated Runs}
Consistency is evaluated for $k=3$ repeated executions of the same experimental condition. Consistency is assessed using cross-run field coverage, presence stability, and value consistency. 

For field $i$, cross-run coverage $D_i$ is

\vspace{-5mm}
\begin{equation}
\label{eq:crossrun-coverage}
D_i = \frac{1}{k}\sum_{j=1}^{k}\delta\!\left(\hat{v}_i^{(j)}\right),
\end{equation}

\noindent where $\delta(\hat{v}_i^{(j)})$ is $1$ if field $i$ is populated with a non-null value in run $j$ and 0 otherwise.

Presence stability $S_i$ is

\vspace{-5mm}
\begin{equation}
\label{eq:presence-stability}
S_i =
\begin{cases}
1, & D_i=0\text{ or }D_i=1,\\
0, & \text{otherwise.}
\end{cases}
\end{equation}

A field that is absent in all runs is therefore stable in its absence, but has zero cross-run coverage.
Value consistency is evaluated among runs in which field $i$ is complete and provides a comparable normalized value. Let $n_i$ denote the number of these runs. Value consistency is defined as

\vspace{-5mm}
\begin{equation}
\label{eq:value-consistency}
V_i = \frac{f_{\mathrm{mode},i}}{n_i},
\qquad n_i \geq 2.
\end{equation}

\noindent Here, $f_{\mathrm{mode},i}$ is the frequency of the most common normalized value. Missing or incomplete fields are not considered when calculating value consistency. If fewer than two comparable runs are available, value consistency is reported as not applicable. The measures describe how consistently the model produces field values across repeated runs, irrespective of their correctness against the ground truth.

\section{Results}\label{sec:results}
\subsection{DPP Generation and Structural AAS Validation}
The experimental evaluation comprises 4,560 generation attempts across 76 product–submodel combinations, five LLMs, and four prompt configurations. Of these, 4,443 outputs (97.4\%) are syntactically valid JSON. A total of 2,205 outputs (48.4\%) passed the AAS 3.2 schema validation, while 2,049 outputs (44.9\%) passed the additional structural validation checks.
The highest full-validator pass rate across all prompt context levels is obtained by GPT-5.6 Sol (91.1\%), followed by Claude Fable 5 (66.9\%) and Qwen 3.5 (53.4\%). Lower pass rates are observed for Mistral Large 4 (12.8\%) and Llama 4 Scout (0.4\%). Table \ref{tab:validation-results} gives a summary of the validation results.

\begin{table}[!ht]
\centering
\caption{Validation results of generated passports across all generated outputs.}
\label{tab:validation-results}
\footnotesize
\setlength{\tabcolsep}{4pt}
\begin{tabular}{@{}lccc@{}}
\toprule
\textbf{Model} &
\shortstack{\textbf{Valid}\\\textbf{JSON}} &
\shortstack{\textbf{AAS}\\\textbf{Schema}} &
\shortstack{\textbf{Full AAS}\\\textbf{Validator}} \\
\midrule
Claude Fable 5 & 100.0\% & 66.9\% & 66.9\% \\
GPT-5.6 Sol & 100.0\% & 91.3\% & 91.1\% \\
Llama 4 Scout & 100.0\% & 6.6\% & 0.4\% \\
Mistral Large 4 & 87.3\% & 15.8\% & 12.8\% \\
Qwen 3.5 & 99.9\% & 61.2\% & 53.4\% \\
\midrule
\textbf{Overall} & \textbf{97.4\%} &
\textbf{48.4\%} & \textbf{44.9\%} \\
\bottomrule
\end{tabular}
\end{table}

The validation results also differ across prompt configurations (see Table \ref{tab:validator-prompts}). GPT-5.6 Sol achieves 79.8\% full AAS validity even under `Low' guidance, suggesting possible prior familiarity with AAS structures. However, this cannot be confirmed without knowledge of its training data. Claude Fable 5 declines from `Medium' context (90.4\%) to `High' context (86.0\%). Qwen 3.5 shows a similar decrease from 76.3\% to 72.4\%. The results suggest that extra guidance may introduce competing instructions and reduce structural compliance. However, the found declines do not establish a causal relationship.

\begin{table}[!ht]
\centering
\caption{Full AAS validator pass rates by LLM and prompt configuration.}
\label{tab:validator-prompts}
\footnotesize
\setlength{\tabcolsep}{4pt}
\begin{tabular}{@{}lccccc@{}}
\toprule
\textbf{Model} & \textbf{Low} & \textbf{Baseline} &
\textbf{Medium} & \textbf{High} & \textbf{Overall} \\
\midrule
Claude Fable 5 & 1.8\% & 89.5\% & 90.4\% & 86.0\% & \textbf{66.9\%} \\
GPT-5.6 Sol & 79.8\% & 96.1\% & 94.3\% & 94.3\% & \textbf{91.1\%} \\
Llama 4 Scout & 0.0\% & 1.8\% & 0.0\% & 0.0\% & \textbf{0.4\%} \\
Mistral Large 4 & 0.9\% & 16.7\% & 15.8\% & 18.0\% & \textbf{12.8\%} \\
Qwen 3.5 & 3.1\% & 61.8\% & 76.3\% & 72.4\% & \textbf{53.4\%} \\
\midrule
\textbf{Overall} & \textbf{17.1\%} & \textbf{53.2\%} &
\textbf{55.4\%} & \textbf{54.1\%} & \textbf{44.9\%} \\
\bottomrule
\end{tabular}
\end{table}

To conclude, the differences between the three validation stages indicate that syntactically valid JSON outputs do not necessarily satisfy the structural requirements of an AAS Environment. Structural validity must therefore be evaluated separately from the successful generation of JSON output.

\setcounter{savedfigure}{\value{figure}}
\addtocounter{figure}{2}
\begin{figure*}[!b]
\centering
\includegraphics[width=\textwidth]{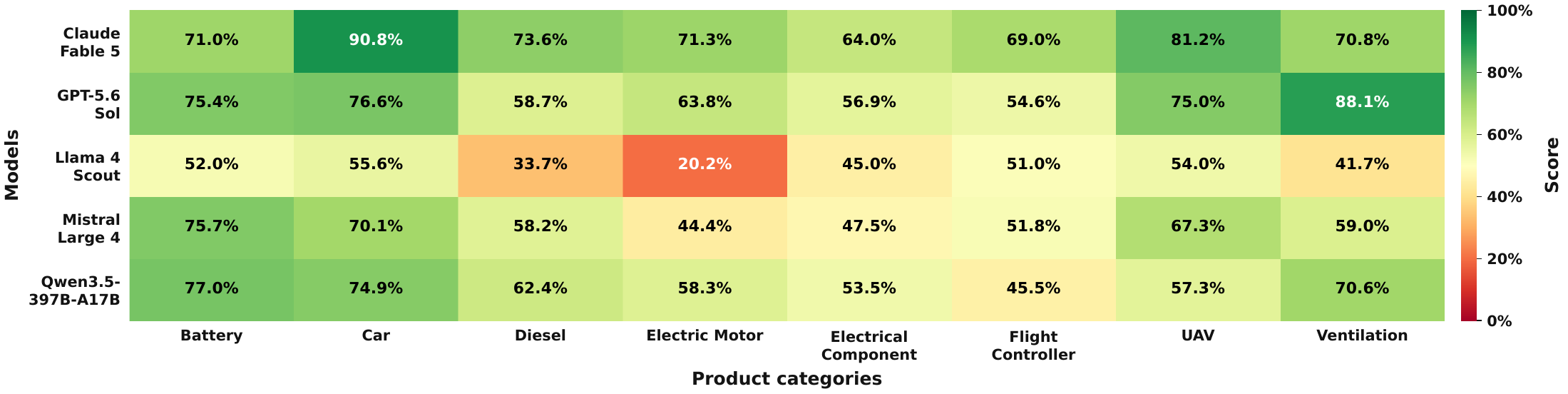}
\caption{Mean conditional accuracy by LLM and industrial product category.}
\label{fig:accuracy-product-category}
\end{figure*}
\setcounter{figure}{\value{savedfigure}}

\subsection{Completeness, Conditional Accuracy and OFC} 
\begin{figure}[H]
\centering
\includegraphics[width=\columnwidth]{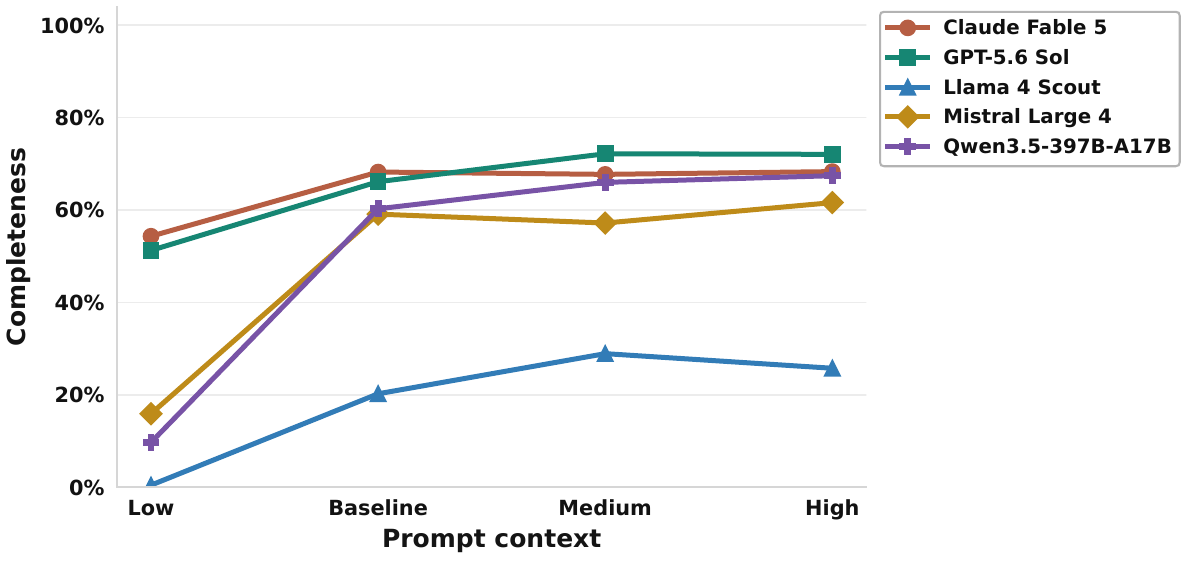}
\caption{Completeness by LLM and prompt context.}
\label{fig:completeness}
\end{figure}

The generated instances are evaluated against the corresponding ground truth AAS environments. Figure \ref{fig:completeness} shows the aggregated completeness scores in \% for the five models and four prompt context configurations. 

Across all experimental conditions, the mean completeness is 49.6\%, showing that approximately half of the required reference fields are populated. Across all generated AAS environments, the mean additional field count is 26 fields per run, representing fields not present in the corresponding ground-truth environments.  \\

\begin{figure}[!ht]
\centering
\includegraphics[width=\columnwidth]{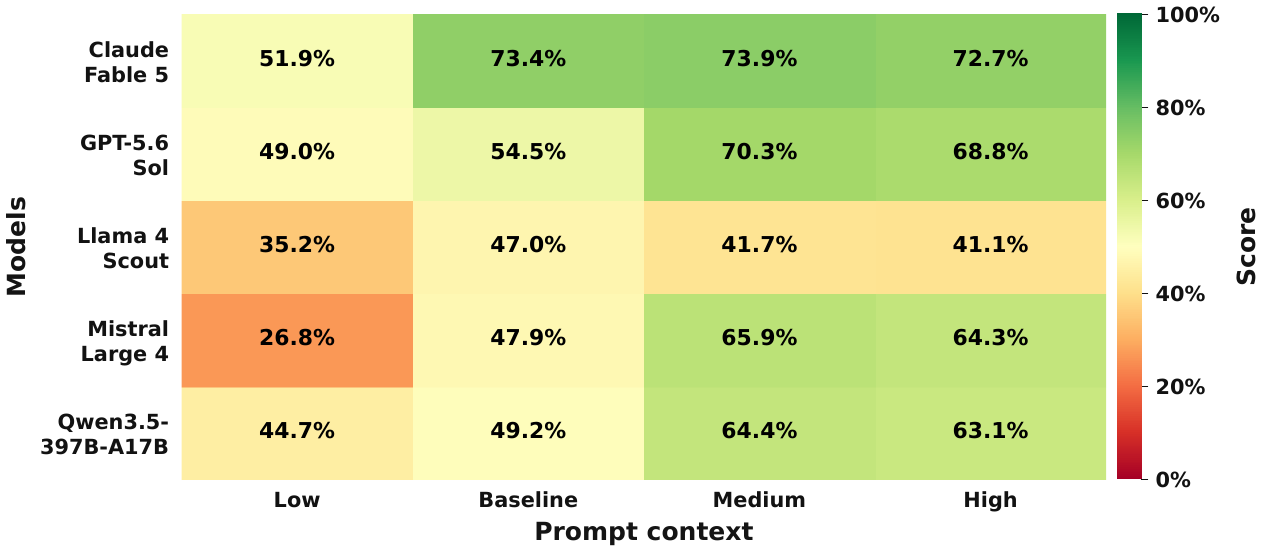}
\caption{Conditional accuracy by LLM and prompt context.}
\label{fig:accuracy}
\end{figure}

Conditional accuracy among complete fields is 58.3\%. This indicates that slightly more than half of the evaluated values are classified as correct by the deterministic scoring procedure. Conditional accuracy results are shown in Figure \ref{fig:accuracy}.

\addtocounter{figure}{1}

The mean OFC is 30.8\% (see Table \ref{tab:ofc}). This lower result reflects the combined influence of missing fields and incorrect or unmatched values, providing a more comprehensive assessment of generation performance than conditional accuracy alone.

\begin{table}[!ht]
\centering
\caption{Mean OFC by LLM and prompt configuration.}
\label{tab:ofc}
\footnotesize
\setlength{\tabcolsep}{4pt}
\begin{tabular}{@{}lccccc@{}}
\toprule
\textbf{Model} & \textbf{Low} & \textbf{Baseline} &
\textbf{Medium} & \textbf{High} & \textbf{Overall} \\
\midrule
Claude Fable 5 & 32.8\% & 52.1\% & 52.2\% & 51.7\% & \textbf{47.2\%} \\
GPT-5.6 Sol & 23.4\% & 34.7\% & 52.6\% & 52.1\% & \textbf{40.7\%} \\
Llama 4 Scout & 0.2\% & 8.9\% & 10.7\% & 9.7\% & \textbf{7.4\%} \\
Mistral Large 4 & 5.4\% & 27.0\% & 39.4\% & 41.5\% & \textbf{28.3\%} \\
Qwen 3.5 & 5.0\% & 28.2\% & 44.9\% & 44.4\% & \textbf{30.6\%} \\
\midrule
\textbf{Overall} & \textbf{13.4\%} & \textbf{30.2\%} &
\textbf{40.0\%} & \textbf{39.9\%} & \textbf{30.8\%} \\
\bottomrule
\end{tabular}
\end{table}

\subsection{Differences Between Submodels and Product Categories}
Conditional accuracy is also compared across eight industrial product categories and the investigated AAS submodels. Figure \ref{fig:accuracy-product-category} shows the results by product category and LLM. Claude Fable 5 achieved the highest conditional accuracy for cars (90.8\%), while GPT-5.6 Sol achieved its highest value for ventilation products (88.1\%). Across the five models, lower values are generally observed for diesel engines, electric motors, and electrical components. Nevertheless, the pattern differs between models as well. For example, Qwen 3.5 achieves 77.0\% for batteries but only 45.6\% for flight controllers.\\

Table \ref{tab:accuracy-submodel-regular} presents conditional accuracy for the general AAS submodel types.
For `Carbon Footprint', the highest values are obtained by GPT-5.6 Sol (99.1\%), Qwen 3.5 (89.1\%), and Claude Fable 5 (91.7\%). The results for `Technical Data' are lower, ranging from 40.3\% for Llama 4 Scout to 57.3\% for Claude Fable 5. 

\begin{table}[!ht]
\centering
\caption{Mean conditional accuracy by LLM and general AAS submodel.}
\label{tab:accuracy-submodel-regular}
\scriptsize
\setlength{\tabcolsep}{3pt}
\begin{tabular}{@{}lccccc@{}}
\toprule
\textbf{Model} & \shortstack{\textbf{Carbon}\\\textbf{Footprint}} & \shortstack{\textbf{Digital}\\\textbf{Nameplate}} & \shortstack{\textbf{Technical}\\\textbf{Data}} & \textbf{Overall} \\
\midrule
Claude Fable 5 & 91.7\% & 74.9\% & 57.3\% & \textbf{67.9\%} \\
GPT-5.6 Sol & 99.1\% & 61.8\% & 52.3\% & \textbf{59.7\%} \\
Llama 4 Scout & 51.9\% & 44.3\% & 40.3\% & \textbf{42.8\%} \\
Mistral Large 4 & 76.4\% & 51.8\% & 48.0\% & \textbf{51.5\%} \\
Qwen 3.5 & 89.1\% & 54.9\% & 53.3\% & \textbf{56.3\%} \\
\midrule
\textbf{Overall} & \textbf{86.1\%} & \textbf{59.1\%} & \textbf{50.8\%} & \textbf{56.9\%} \\
\bottomrule
\end{tabular}
\end{table}

Table \ref{tab:accuracy-submodel-battery} presents conditional accuracy for the battery AAS submodel types. The battery-specific submodels exhibited substantial variation between models. For example, Llama 4 Scout achieved 97.4\% for `Battery Material Composition' but 0.0\% for `Battery Digital Nameplate', while Mistral Large 4 reached 100.0\% for `Battery Product Carbon Footprint'.

\begin{table}[!ht]
\centering
\caption{Mean conditional accuracy by LLM and battery AAS submodel.}
\label{tab:accuracy-submodel-battery}
\scriptsize
\setlength{\tabcolsep}{2pt}
\begin{tabular}{@{}lccccc@{}}
\toprule
\textbf{Model} & \shortstack{\textbf{Battery}\\\textbf{Material}\\\textbf{Composition}} & \shortstack{\textbf{Battery}\\\textbf{Digital}\\\textbf{Nameplate}} & \shortstack{\textbf{Battery}\\\textbf{Product}\\\textbf{Carbon}\\\textbf{Footprint}} & \shortstack{\textbf{Battery}\\\textbf{Technical}\\\textbf{Data}} & \textbf{Overall} \\
\midrule
Claude Fable 5 & 82.0\% & 95.0\% & 33.3\% & 54.8\% & \textbf{71.0\%} \\
GPT-5.6 Sol & 77.2\% & 77.3\% & 72.2\% & 73.3\% & \textbf{75.4\%} \\
Llama 4 Scout & 97.4\% & 0.0\% & 33.3\% & 42.0\% & \textbf{52.0\%} \\
Mistral Large 4 & 71.7\% & 77.1\% & 100.0\% & 66.2\% & \textbf{75.7\%} \\
Qwen 3.5 & 92.0\% & 80.3\% & 75.0\% & 59.7\% & \textbf{77.0\%} \\
\midrule
\textbf{Overall} & \textbf{82.8\%} & \textbf{76.5\%} & \textbf{62.5\%} & \textbf{59.8\%} & \textbf{71.5\%} \\
\bottomrule
\end{tabular}
\end{table}

\subsection{Consistency Across Repeated Runs}\label{sec:consistency}
Consistency is assessed using cross-run field coverage, presence stability, and value consistency. The analysis identifies fields that are consistently populated, fields that are repeatedly omitted, and fields for which generated values varied across repeated executions. These results complement the field-correctness evaluation, as repeated generation of the same value does not necessarily mean that the value is correct.\\

Mean cross-run field coverage is 49.6\%, with required fields populated in approximately half of the repeated executions. GPT-5.6 Sol achieves the highest average coverage (65.4\%), followed by Claude Fable 5 (64.6\%), while Llama 4 Scout shows substantially lower coverage (18.8\%). Table \ref{tab:cross-run} summarizes the cross-run field coverage results.

\begin{table}[!ht]
\centering
\caption{Mean cross-run field coverage by LLM and prompt configuration.}
\label{tab:cross-run}
\footnotesize
\setlength{\tabcolsep}{4pt}
\begin{tabular}{@{}lccccc@{}}
\toprule
\textbf{Model} & \textbf{Low} & \textbf{Baseline} &
\textbf{Medium} & \textbf{High} & \textbf{Overall} \\
\midrule
Claude Fable 5 & 54.3\% & 68.2\% & 67.7\% & 68.3\% & \textbf{64.6\%} \\
GPT-5.6 Sol & 51.2\% & 66.1\% & 72.2\% & 72.0\% & \textbf{65.4\%} \\
Llama 4 Scout & 0.4\% & 20.2\% & 28.9\% & 25.7\% & \textbf{18.8\%} \\
Mistral Large 4 & 15.9\% & 59.1\% & 57.2\% & 61.6\% & \textbf{48.5\%} \\
Qwen 3.5 & 9.7\% & 60.3\% & 66.0\% & 67.4\% & \textbf{50.8\%} \\
\midrule
\textbf{Overall} & \textbf{26.3\%} & \textbf{54.8\%} &
\textbf{58.4\%} & \textbf{59.0\%} & \textbf{49.6\%} \\
\bottomrule
\end{tabular}
\end{table}

Mean field presence stability is 90.5\% across all experimental conditions, indicating that most required fields are either consistently populated or consistently omitted across the three runs. The highest stability is observed for Claude Fable 5 (95.5\%) and Llama 4 Scout (95.1\%) (see Table \ref{tab:field-stability}). However, high stability does not mean successful information extraction.

\begin{table}[!ht]
\centering
\caption{Field stability by LLM and prompt configuration.}
\label{tab:field-stability}
\footnotesize
\setlength{\tabcolsep}{4pt}
\begin{tabular}{@{}lccccc@{}}
\toprule
\textbf{Model} & \textbf{Low} & \textbf{Baseline} &
\textbf{Medium} & \textbf{High} & \textbf{Overall} \\
\midrule
Claude Fable 5 & 95.8\% & 96.4\% & 95.7\% & 94.1\% & \textbf{95.5\%} \\
GPT-5.6 Sol & 85.1\% & 87.6\% & 93.9\% & 93.1\% & \textbf{89.9\%} \\
Llama 4 Scout & 99.0\% & 93.5\% & 93.2\% & 94.6\% & \textbf{95.1\%} \\
Mistral Large 4 & 79.7\% & 85.5\% & 70.7\% & 78.1\% & \textbf{78.5\%} \\
Qwen 3.5 & 89.1\% & 93.9\% & 95.1\% & 95.6\% & \textbf{93.4\%} \\
\midrule
\textbf{Overall} & \textbf{89.7\%} & \textbf{91.4\%} &
\textbf{89.7\%} & \textbf{91.1\%} & \textbf{90.5\%} \\
\bottomrule
\end{tabular}
\end{table}

\begin{figure}[!ht]
\centering
\includegraphics[width=\columnwidth]{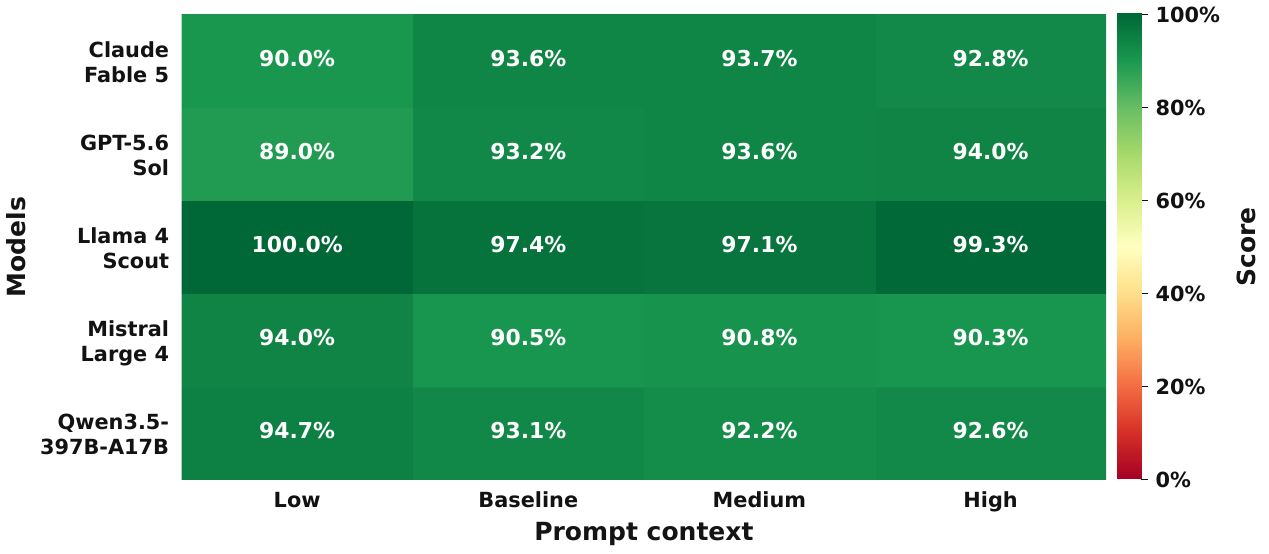}
\caption{Value consistency by LLM and prompt context.}
\label{fig:consistency}
\end{figure}

Mean value consistency is 93.1\% across eligible experimental conditions (see Figure \ref{fig:consistency}).

Scores range from 89.0\% to 100.0\%. This indicates that comparable populated fields generally contained the same values across repeated executions. Llama 4 Scout achieved the highest consistency across all prompt configurations. The result does not necessarily indicate high completeness or accuracy.

\subsection{Deterministic and LLM-as-a-Judge Scoring Differences}
\begin{figure}[!ht]
\centering
\includegraphics[width=\columnwidth]{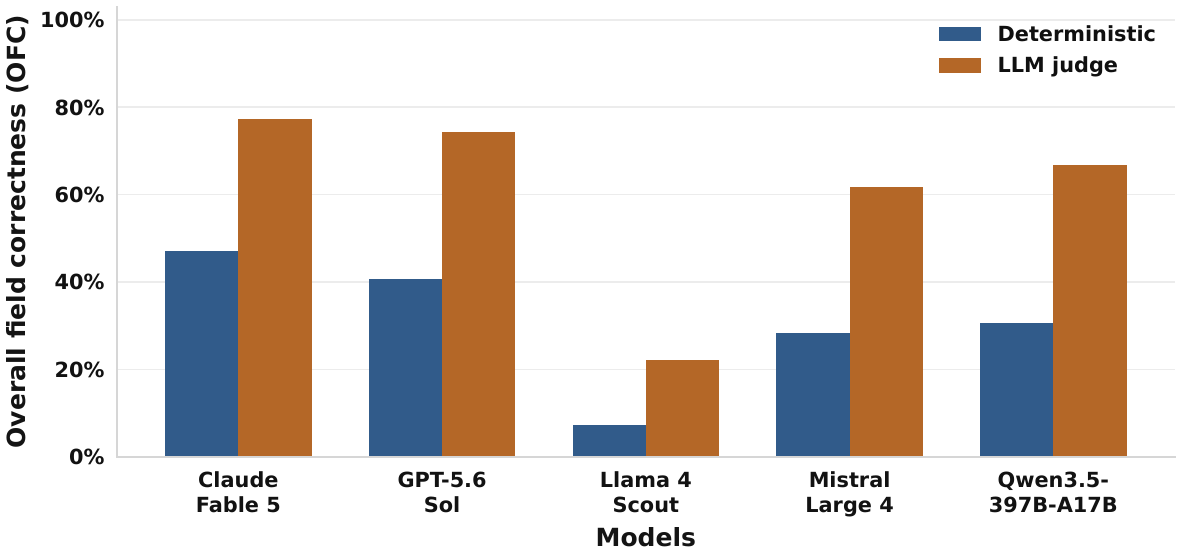}
\caption{OFC Deterministic and LLM-as-a-Judge.}
\label{fig:ofc-comparison}
\end{figure}

The LLM-as-a-judge procedure returned higher overall completeness (66.5\%, compared with 49.6\% under deterministic scoring), conditional accuracy (88.9\% versus 58.3\%), and OFC (60.4\% versus 30.8\%). The OFC difference is 29.6 percentage points for the same 4,560 generation attempts (also see Figure \ref{fig:ofc-comparison}). In the paired run-level comparison, judge-based OFC is higher in 3,807 cases, equal in 744, and lower in nine. The scores therefore reflect substantial differences between the evaluators rather than changes to the generated AAS instances. Strict deterministic field matching may underestimate extraction quality (semantically equivalent properties may be rejected). However, without further human validation, the higher LLM-as-a-judge scores cannot be considered more accurate. 

The retrieval evaluation achieved an evidence recall of 98\%, indicating that most source-supported ground-truth values are represented in the retrieved context. 

\subsection{Prompt Reduction and Token Consumption}\label{sec:token}
\begin{figure}[!ht]
\centering
\includegraphics[width=\columnwidth]{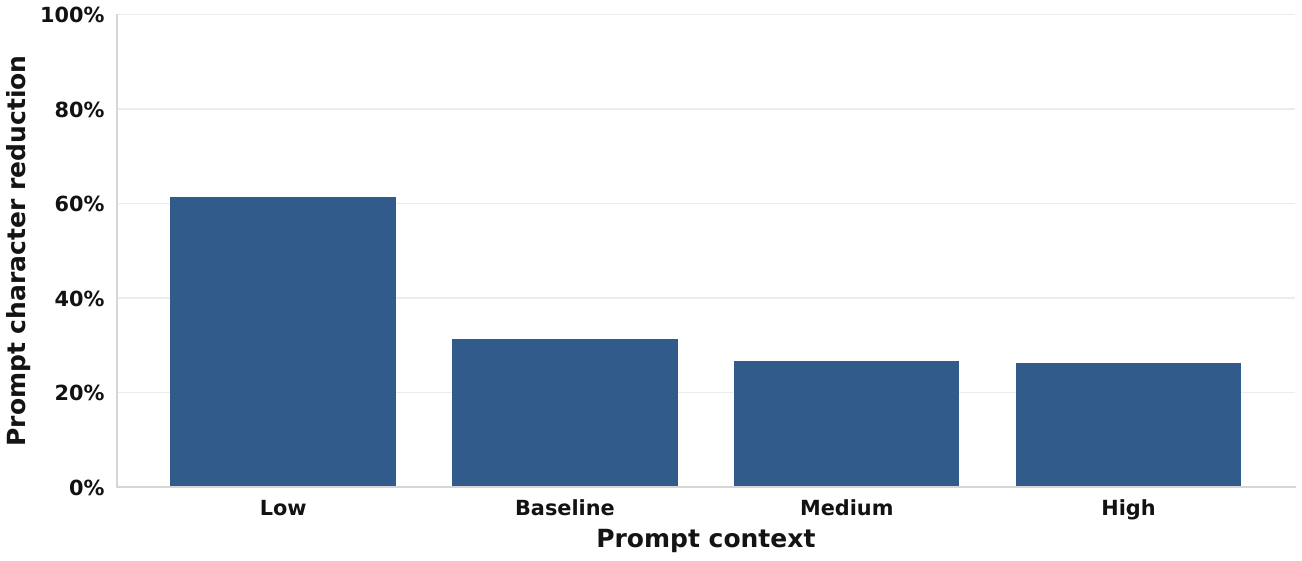}
\caption{Weighted mean prompt character reduction by prompt context.}
\label{fig:prompt-reduction}
\end{figure}

The requirement-driven RAG approach reduces prompt length by selecting relevant information instead of providing the complete source documentation (see Figure \ref{fig:prompt-reduction}). 

The weighted mean prompt character reduction ranged from 26.2\% for `High' to 61.4\% for `Low' guidance. `Baseline' and `Medium' achieved reductions of 31.4\% and 26.6\%, respectively. The product documentation used in the given study largely fit within the context windows of the evaluated LLMs. However, industrial applications involving extensive product documentation and life cycle data may exceed these limits. Requirement-driven RAG therefore provides a scalable approach by restricting the supplied context to information relevant to the target AAS submodel. By reducing the amount of supplied information, RAG can help lower token consumption and associated inference costs.\\

\begin{figure}[!ht]
\centering
\includegraphics[width=\columnwidth]{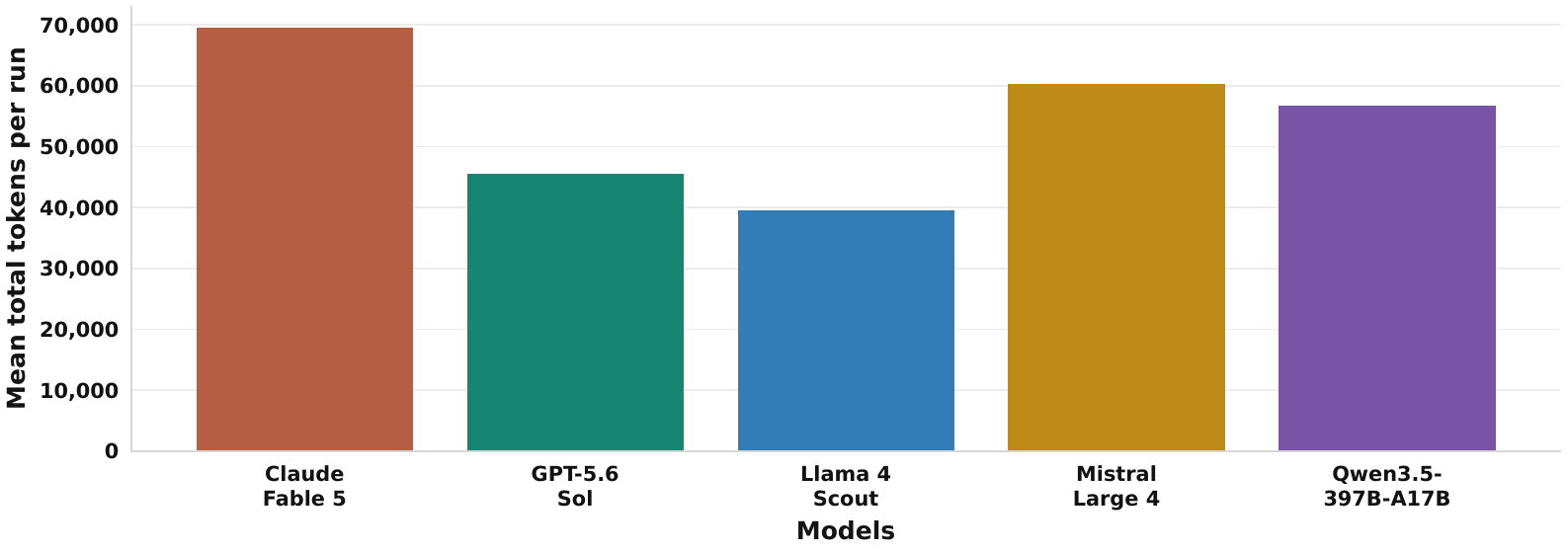}
\caption{Mean total tokens per run by LLM.}
\label{fig:tokens}
\end{figure}

The token analysis additionally reveals differences between LLMs, with Claude Fable 5 consuming approximately 69,000 total tokens per execution, compared with approximately 39,000 for Llama 4 Scout (see Figure \ref{fig:tokens}).

\section{Discussion}\label{sec:discussion}
\subsection{Automated DPP Creation in a Circular Economy}
The experimental results provide evidence concerning the extent to which existing industrial documentation can be used to generate standardized AAS-based product information. The proposed approach reduces the dependence on manual extraction and mapping by using standardized submodel requirements to guide information retrieval and generation. It is therefore applicable to organizations that possess product documentation but have not yet established repositories of populated AAS instances.\\

However, automatically generated AAS instances are only one part of a functioning DPP infrastructure. Their use in circular manufacturing also requires reliable product identification, data traceability, access control, and life cycle information management.

\subsection{Product Data Origins}
Product information can originate from various systems and have heterogeneous formats. The implemented pipeline retains information about retrieved source documents and document chunks, but does not establish complete traceability to the authoritative information source. For industrial deployment, additional mechanisms would be needed to maintain relationships between generated AAS values, original source records, product revisions, and subsequent life cycle updates.

\subsection{Limitations}
First, the evaluation is limited to the selected industrial products, source documents, and standardized AAS submodels. Despite covering eight product categories, the unbalanced dataset limits the generalizability of the findings. Additionally, the completeness and accuracy metrics depend on the correctness of the manually created ground truths.

Second, the pipeline only processes static documents. The experiments consequently do not assess live data exchange, continuous passport updates, or the handling of life cycle information generated during manufacturing.

Third, the evaluation methods used differ. Deterministic scoring is reproducible but may fail to recognize semantically equivalent values. An LLM-as-a-judge allows for more flexible comparisons but may introduce less reproducible judgments.

Finally, the study does not establish whether requirement-driven RAG outperforms alternative retrieval approaches.

\section{Conclusion}\label{sec:conclusion}
The paper showed a pipeline to automatically generate AAS-based DPPs and DBPs with the help of LLMs. The results provide a basis for assessing the suitability of the investigated LLMs for preparing standardized product passports.\\

Across 4,560 generations, 97.4\% of responses are valid JSON, but only 44.9\% passed the full AAS validator. Under deterministic scoring, mean completeness, conditional accuracy, and OFC are 49.6\%, 58.3\%, and 30.8\%, respectively. The corresponding LLM-as-a-judge scores are 66.5\%, 88.9\%, and 60.4\%. The 29.6 percentage-point difference in mean OFC between deterministic and LLM-based scoring shows that evaluation design influences reported performance.\\

Four prompt context levels show differences in the creation performance of the passports under investigation. Deterministic OFC increases from 13.4\% for Low guidance to 30.2\% for `Baseline' and approximately 40\% at `Medium' and `High', indicating that providing the AAS template is associated with better field-level results without establishing statistical significance. Claude Fable 5 achieves the highest deterministic OFC (47.2\%), while GPT-5.6 Sol achieves the highest full AAS validator pass rate (91.1\%). Field presence stability averages 90.5\%, but is also increased by consistently absent fields. The value consistency score averages 93.1\%.\\

For circular manufacturing, the pipeline can help turn scattered product data into standardized DPPs that inform component reuse and material recovery. The use of standardized AAS submodels gives product stakeholders an interoperable product information exchange format across life cycles.\\

Future work should use more balanced datasets and link DPP fields to authoritative product records. These links can aid LLM-assisted DPP updates after products leave a factory.



\newpage
\appendix
\section{Supplementary Research Materials}\label{sec:materials}

\noindent The supplementary research materials are available at:\\

\noindent \href{https://github.com/mbuecheler/automated-LLM-generated-DPPs}{%
\texttt{https://github.com/mbuecheler/}\\
\texttt{automated-LLM-generated-DPPs}}



\begingroup
\raggedright
\interlinepenalty=10000
\bibliographystyle{elsarticle-num-accessed}
\bibliography{references}
\endgroup

\end{document}